\documentclass[superscriptaddress,twocolumn,pra]{revtex4-1}
\usepackage[export]{adjustbox}

\usepackage{wrapfig}
\usepackage{float}
\usepackage{times}
\usepackage{color}

\usepackage{braket}
\usepackage{graphicx} 
\usepackage[space]{grffile}

\usepackage{amsmath}
\usepackage{amssymb}

\usepackage{accents}
\usepackage{datetime}
\usepackage{longtable}
\usepackage{amsmath}
\usepackage{longtable}
\usepackage{amssymb}
\usepackage{latexsym}
\usepackage{color}
\usepackage{bm}
\usepackage{mathtools}

\usepackage{times}
\usepackage{epsfig}
\usepackage{wrapfig}
\usepackage[T1]{fontenc}
\usepackage{float}
\usepackage{enumerate}
\usepackage[compact]{titlesec}
\usepackage{setspace,ifthen}
\usepackage{amsthm} 
\usepackage{amssymb}	
\usepackage{times}
\usepackage[usenames,dvipsnames]{xcolor}
\usepackage{mathrsfs,amsmath}
\usepackage{mathrsfs}
\usepackage{mathrsfs}
\usepackage{rotating}
\usepackage{epstopdf}
\usepackage{appendix}

\usepackage{csquotes} 
\usepackage{bbold} 
\usepackage{datetime} 
\usepackage{units} 
\usepackage{afterpage}

\usepackage{array}
\usepackage{graphicx}
\usepackage{amsmath}
\usepackage{longtable}
\usepackage{amssymb}
\usepackage{latexsym}
\usepackage{times}
\usepackage{epsfig}
\usepackage{wrapfig}
\usepackage[T1]{fontenc}
\usepackage{float}
\usepackage{enumerate}
\usepackage{setspace,ifthen}
\usepackage{amsthm} 
\usepackage{amssymb}	
\usepackage[usenames,dvipsnames]{xcolor}
\usepackage{mathrsfs,amsmath}
\usepackage{mathrsfs}
\usepackage{mathrsfs}
\usepackage{rotating}
\usepackage{epstopdf}
\usepackage{appendix}
\usepackage{blkarray}
\usepackage{multirow}
\usepackage{cancel}
\usepackage{comment}

\usepackage{array}

\usepackage{rotate}
\usepackage{kbordermatrix} 

\usepackage{rotate}

\usepackage[colorlinks, linkcolor = black, citecolor = black, filecolor = black, urlcolor = blue]{hyperref}

\renewcommand{\kbldelim}{(} 
\renewcommand{\kbrdelim}{)}

\newcommand{\X}{\hat{X}}
\newcommand{\Y}{\hat{Y}}

\newcommand{\sigz}{\hat{\sigma}_z}

\newcommand{\RBseqClean}{\mathcal{S}}
\newcommand{\RBseqNoisy}{\tilde{\mathcal{S}}}

\newcommand{\Identity}{\mathbb{I}}

\newcommand{\Covariance}{\text{Cov}}

\newcommand{\E}[2][1]{\mathbb{E}[#2]}
\newcommand{\Var}[2][1]{\mathbb{V}[#2]}

\newcommand{\Cov}[2][1]{\Covariance\left[#2\right]}

\newcommand{\norm}[1]{\| #1 \|} 

\newcommand{\eqn}[1]{Eq.~\ref{#1}}

\newcommand{\vect}[1]{\accentset{\rightharpoonup}{#1}}

\newcommand{\sigLSq}{\sigma_\textrm{L}^2}
\newcommand{\sigSSq}{\sigma_\textrm{S}^2}
\newcommand{\Me}{\mathcal{M}_\varepsilon}
\newcommand{\Mn}{\mathcal{M}_\mathrm{n}}

\newcommand{\dS}{\delta_\textrm{S}}
\newcommand{\dL}{\delta_\textrm{L}}

\newcommand{\sigx}{\hat{\sigma}_x}
\newcommand{\sigy}{\hat{\sigma}_y}
\newcommand{\errorVect}{\vec{\textbf{a}}}
\newcommand{\noiseAve}[1]{\langle {#1} \rangle_n}
\newcommand{\RvecL}{\vec{\boldsymbol{R}}_\textrm{L}}
\newcommand{\VvecL}{\vec{\boldsymbol{V}}_\textrm{L}}
\newcommand{\RvecS}{\vec{\boldsymbol{R}}_\textrm{S}}
\newcommand{\RvecLXY}{\vec{\boldsymbol{R}}_\textrm{L,2D}}
\newcommand{\RvecSXY}{\vec{\boldsymbol{R}}_\textrm{S,2D}}
\newcommand{\VvecLXY}{\vec{\boldsymbol{V}}_\textrm{L,2D}}
\newcommand{\RnormSQSXY}{\norm{\RvecSXY}^2}
\newcommand{\VnormSQLXY}{\norm{\VvecLXY}^2}
\newcommand{\sigLFour}{\sigma_\textrm{L}^4}
\newcommand{\sigSFour}{\sigma_\textrm{S}^4}

\begin{document}

\title{Measuring and Suppressing Error Correlations in Quantum Circuits}
\author{C. L. Edmunds}
\affiliation{ARC Centre for Engineered Quantum Systems, School of Physics, The University of Sydney, NSW Australia}
\affiliation{National Measurement Institute, West Lindfield NSW 2070 Australia}
\author{C. Hempel}
\affiliation{ARC Centre for Engineered Quantum Systems, School of Physics, The University of Sydney, NSW Australia}
\affiliation{National Measurement Institute, West Lindfield NSW 2070 Australia}
\author{R. Harris}
\affiliation{ARC Centre for Engineered Quantum Systems, School of Physics and Mathematics, The University of Queensland, St Lucia, QLD Australia}
\author{H. Ball}
\affiliation{ARC Centre for Engineered Quantum Systems, School of Physics, The University of Sydney, NSW Australia}
\author{V. Frey}
\affiliation{ARC Centre for Engineered Quantum Systems, School of Physics, The University of Sydney, NSW Australia}
\author{T. M. Stace}
\affiliation{ARC Centre for Engineered Quantum Systems, School of Physics and Mathematics, The University of Queensland, St Lucia, QLD Australia}
\author{M. J. Biercuk$ ^{\dagger} $}
\affiliation{ARC Centre for Engineered Quantum Systems, School of Physics, The University of Sydney, NSW Australia}
\affiliation{National Measurement Institute, West Lindfield NSW 2070 Australia}
\email{ $^{\dagger}$Contact: michael.biercuk@sydney.edu.au }
\date{\today~at~\currenttime}

\begin{abstract}
Quantum error correction provides a path to large-scale quantum computers, but is built on challenging assumptions about the characteristics of the underlying errors.  In particular, the mathematical assumption of independent errors in quantum logic operations is at odds with realistic environments where error-sources may exhibit strong temporal correlations.  We present experiments enabling the identification of error correlations between operations in quantum circuits, using only projective measurements at the end of the circuit.  Using a single trapped ion qubit and engineered noise with tunable temporal correlations, we identify a clear signature of error correlations between sequential gates in randomly composed quantum circuits, and extract quantitative measures linked to the underlying noise correlation length. By replacing all gates in these circuits with ``virtual'' dynamically corrected gates (DCGs), we demonstrate that even in the presence of strongly correlated noise the signatures of error correlations between sequential gates appear similar to standard gates exposed to uncorrelated noise.  A theoretical model applied to our experiments reveals that common DCGs suppress the correlated error component by over $270\times$ with $95\%$ confidence.  Using block-correlated noise, we explore the scaling of the effective error correlation length at the virtual level, and show that DCGs exhibit error correlations indistinguishable from those arising from uncorrelated noise. 
\end{abstract}

\maketitle

Suppressing and correcting errors in quantum circuits is a critical challenge driving a substantial fraction of research in the quantum information science community.   These efforts build on quantum error correction (QEC) and the theory of fault tolerance as the fundamental developments that support the concept of large-scale quantum computation~\cite{PreskillQEC,NC2000, QECLidar2013}.  
In combination, these theoretical constructs suggest that so long as the probability of error in each physical quantum information carrier
can be reduced below a threshold value, a properly executed QEC protocol can detect and suppress logical errors to arbitrarily low levels, and hence enable arbitrarily large computations.  
Underlying this proposition is an assumption that errors are statistically independent, {\em i.e.} the emergence of a qubit error at a specific time is uncorrelated with errors arising in other qubits or at any other times~\cite{Preskill_correlations2013}.

The practicality of this assumption has long been questioned, as laboratory sources of noise commonly exhibit strong temporal correlations, captured through spectral measures exhibiting high weight at low frequencies. Attempts to address this in the theory of quantum error correction are challenging and results to date suggest that revision of postulated fault-tolerant thresholds may be required~\cite{Preskill2009, Preskill2006} relative to more optimistic predictions that have recently emerged~\cite{Fowler_Surface}.  Without the development of new techniques to suppress statistical correlations in the underlying errors, the future efficacy of QEC thus depends on suitable, but likely complex, reanalysis and modification of correction routines~\cite{Preskill_correlations2013}.  Here we turn to a technique tied to a higher abstraction layer than the physical gates; replacing physical qubit operations with logically equivalent dynamically corrected gates (DCGs) forms a virtual layer wherein error correlations can in principle be reduced before the application of QEC~\cite{Preskill_Layered, JonesPRX2012}, augmenting laboratory-based technical approaches to minimize noise.  Still, it remains a significant technical challenge to even  measure error correlations in quantum circuits when projective measurement collapses quantum states.

In this work we experimentally demonstrate techniques to detect and suppress temporal correlations between errors arising in quantum circuits applied to a trapped-ion qubit. We employ a recently developed analytic framework~\cite{Ball:2016, Mavadia:2017} to identify signatures of correlations between errors in sequential gates using only single-qubit projective measurements performed at the end of randomly composed circuits of Clifford operations~\cite{Emerson2005, Knill2008, Wallman:2016}.  Experiments with a single trapped ion in the presence of engineered noise processes with long or short correlation lengths relative to individual operations validate the predictions of this model. Further investigations replace bare physical gates with error-suppressing DCGs~\cite{Brown2004, Khodjasteh2009dcg, True, DasSarmaGate, SoareNatPhys2014}, and show that DCGs can not only reduce net error rates but also suppress correlations between sequential gate errors by \emph{whitening} the effective error spectrum at the virtual gate layer~\cite{SoareNatPhys2014, ViolaFFF}.  Quantitative analysis allows extraction of residual error contributions and demonstrates over two orders of magnitude reduction in the correlated error component. These results provide direct and strong evidence that the use of dynamically protected physical qubit operations in a layered architecture for quantum computing~\cite{JonesPRX2012} can facilitate the successful application of extant QEC theory with minimal revision on the path to fault-tolerant quantum computation.

\begin{figure*}[ht]
	\label{fig:1}
	\centering
\includegraphics[scale =1]{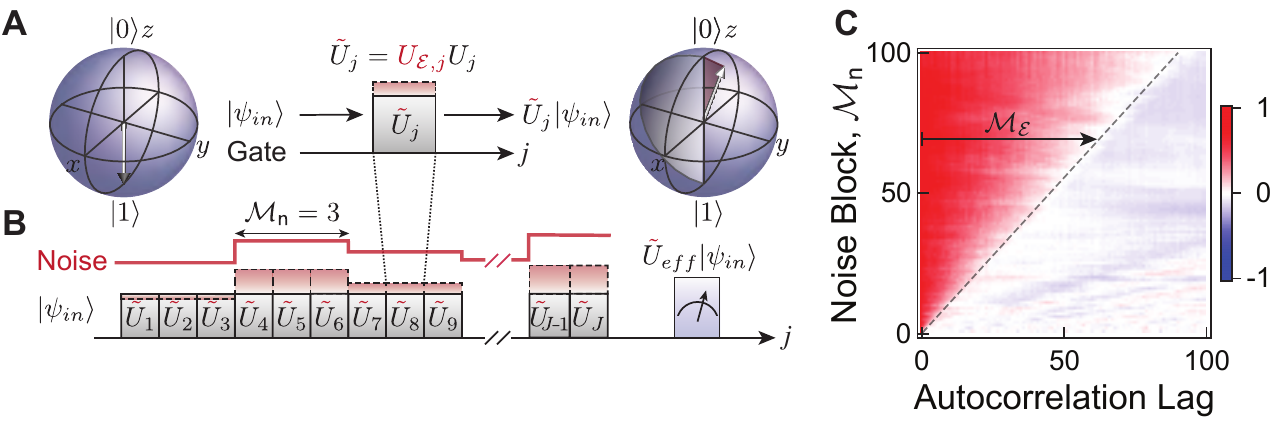}
	\caption{\textbf{Translation of noise to errors in quantum circuits}
(A) A single operation applied to a qubit in the presence of noise, $\tilde{U}_j$, can be decomposed into an error operator $U_{\varepsilon,j}$ and the target operation $U_{j}$. Bloch spheres schematically illustrate the effect of noise on a $X_{\pi}$ gate acting on input state $\ket{1}$, with red shading indicating the resultant error. (B) Noise (red line) exhibiting nonzero temporal correlation of length $\Mn$, quantized in units of gate operations, acts on a quantum circuit composed of sequentially applied unitary operations. The resultant errors accumulate and lead to a noisy effective operator $\tilde{U}_\mathrm{eff}$, whose effect is determined through a projective measurement at the end of the circuit. 
(C) Translation of correlations in the noise process to correlations in the magnitude of the circuit error vector. The error vector for each gate of a randomly composed circuit of 1000 gates under a noise process with correlation noise length $\Mn$ is calculated and the autocorrelation function of the magnitude of the error vector is shown for the first 100 gates.
}

\label{Fig:F1}
\end{figure*}

Various metrics and benchmarking protocols may be employed to infer the behavior of gate operations~\cite{Poyatos:1997, Emerson2005, Knill2008, MagesanInterleaved, Emerson2011, Merkel:2013, Kimmel:2014, Wallman:2015, Sheldon:2016, BlumeKohout:2017, Mavadia:2017} using projective measurements. Here most experimental routines extract the average difference between a qubit state transformed under an imperfect operation, $\tilde{U}$, and a predetermined target state $\ket{\psi_\mathrm{in}}\to\tilde{U}\ket{\psi_\mathrm{in}}$ (Fig.~\ref{Fig:F1}A). In a quantum circuit consisting of $J$ gates, this corresponds to the resulting net noisy state transformation $\tilde{U}_\mathrm{eff}\ket{\psi_\mathrm{in}}$ (Fig.~\ref{Fig:F1}B), which is determined by a complex interplay of both the sensitivity of each individual gate to the noise and the impact of the circuit structure on error accumulation~\cite{GreenNJP2013, Ball:2016, Wallman:2016, Mavadia:2017}.

To obtain an experimentally accessible measure for the effect of correlations, we first investigate how noise correlations of a block length $\Mn$ (Fig.~\ref{Fig:F1}B), in units of logical gates, translate to a correlation length $\Me$ throughout the resultant errors.  Using the filter-transfer function framework~\cite{Kofman2004, GreenPRL2012, GreenNJP2013, ViolaFFF}, the error component in each gate can be written as 
\mbox{$\tilde{U}_{\varepsilon,j}=\exp\left\{{\sum_{\alpha=1}^{\infty}[\vect{\textbf{a}}_j]_{\alpha}\cdot\vect{\sigma}}\right\}$}, 
with Pauli matrices $\vec{\sigma}$, and $\alpha$ an index denoting the Magnus order of expansion~\cite{GreenNJP2013}. 

The relationship between noise correlations and error correlations is revealed by numerically calculating the error vector $\vect{\textbf{a}}_j$ for each operation in a randomly composed single-qubit circuit exposed to noise with varying $\Mn$. Here a circuit is assembled from the 24 Clifford operations comprising combinations of $\pi$ and $\pi/2$ rotations about the $x,y$ and $z$-axes of the Bloch sphere, and an identity gate $\hat{\mathbb{I}}$. The error vector characterizes the strength and nature of the error experienced by each gate; calculation of the autocorrelation function of the magnitude of the error vector across the first 100 gates of a randomly composed 1000-gate circuit reveals strong correlations of length $\Me$ that scale linearly with the correlation length of the input noise process $\Mn$ (Fig.~\ref{Fig:F1}C).

Probing this behavior experimentally is complicated by the nominal inaccessibility of $\vect{\textbf{a}}_j$ when gate $\tilde{U}_{\varepsilon,j}$ is embedded in a circuit, as state-collapse under projective measurement limits us to accessing only the net effective error at the end of the circuit.  We therefore aim to directly link measurement outcomes on single-qubit circuits to the nature of the underlying error correlations quantified by $\Me$. Our approach involves a statistical model largely following the formalism of reference~\cite{Ball:2016}. The latter maps the probability of error on each individual gate in a $J$-gate circuit to a single, unit-length step in the space of Pauli error operators, $\hat{\sigma}_{x,y,z}$.
The length of the resultant $J$-step walk is an intrinsic property of the circuit, and acts as a proxy for its susceptibility to correlated errors.  Examining individual randomly composed circuits that ideally implement the identity operator reveals the idiosyncratic nature of their walks; certain randomly composed circuits exhibit long walks, while others have walks that terminate near the origin.  This general framework linking the Pauli walk to accumulated error was recently experimentally validated in~\cite{Mavadia:2017}.

\begin{figure}[bp]
\label{fig:2}
\centering
\includegraphics[width=8.3cm]{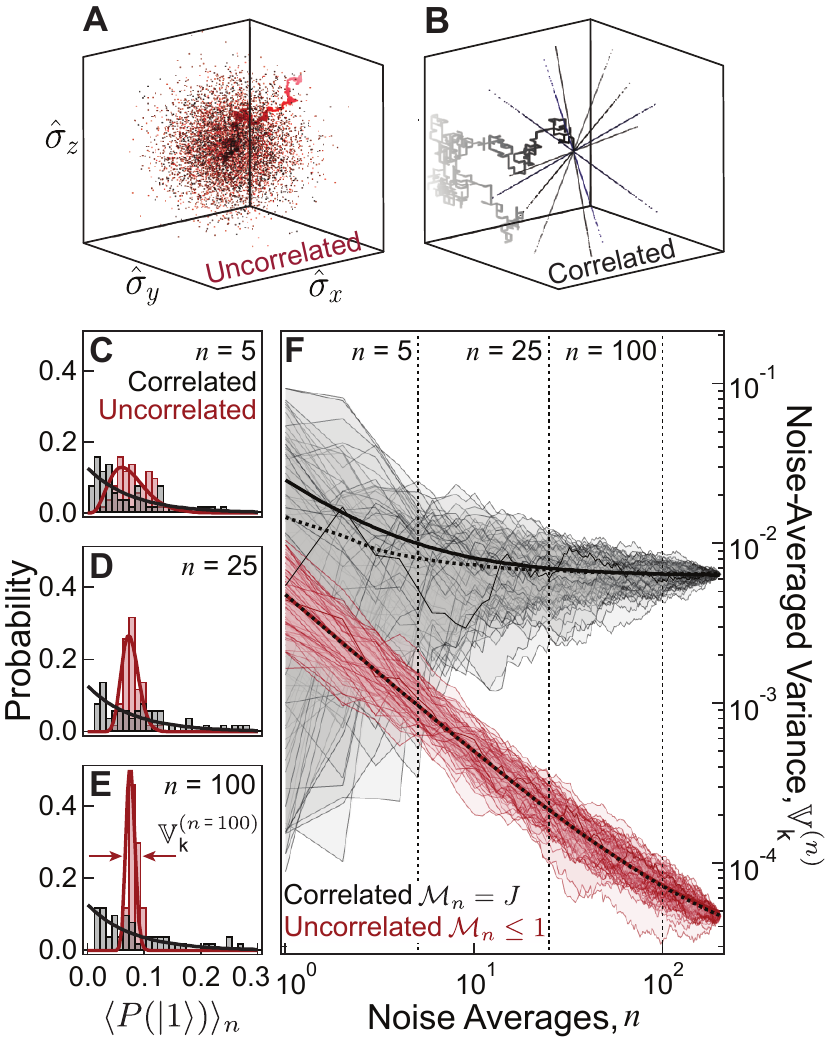}
\caption{\textbf{Signatures of error correlations in randomly composed circuits.} 
(A-B) Random walks for the extremal error-correlation cases, $\Mn \leq 1$ and $\Mn = J$.
Final walk displacements of eight circuits under 1000 error realizations shown along with the full walk for a single circuit that is common between the two error models.  
(C-E) Distribution of measured probabilities for $k=50$ randomly composed circuits averaged over $n=5, 25$ and 100 noise realizations drawn from \mbox{$\delta\sim\mathcal{N}(0,\sigma^2=2\times 10^{-3})$} for both maximally correlated, $\Mn = J$, (gray) and uncorrelated, $\Mn \leq 1$, (red) noise processes. Solid lines are renormalized gamma distributions plotted with no free parameters. Each measured probability is derived from $r=220$ repetitions under identical circuit and noise settings to minimize the impact of quantum projection noise. (F) Scaling of cumulatively noise-averaged histogram variances $\mathbb{V}_\mathrm{k}^{(n)}  \equiv  \mathbb{V}_\mathrm{k}\left[\left\langle P(\ket{1})\right\rangle_{n}\right]$. Trajectories correspond to different orderings of noise realizations with dotted lines representing the mean of 1000 reorderings.  An individual trajectory for each data set is highlighted for clarity.  Vertical dashed lines indicate the values of $n$ used in panels (C)-(E).  Solid lines are fits to the data (see main text).  
}
\label{Fig:F2}
\end{figure}

The experimentally measurable net error at the conclusion of the circuit is obtained by rescaling each step of the walk by the magnitude of the individual gate error, $\tilde{U}_{\varepsilon,j}$, present during that gate. It is the interplay of circuit-specific features with the underlying correlations in the noise that is central to our ability to identify the character of error correlations between individual gates. In the presence of temporally uncorrelated errors, the walk is randomized stepwise, and the influence of circuit structure on net error accumulation is washed out (Fig.~\ref{Fig:F2}A).  Here, the calculated locus of walk termination points for different circuits and error realizations appears randomly distributed in Pauli space, meaning that averaging over experiments would result in a spread of outcomes that grows narrower as the experiment number increases.  By contrast, in the presence of errors with maximal correlation across all gates, the terminations of the resultant walks for the same circuits now appear to be dominated by the underlying circuit structures (``rays'' in Fig.~\ref{Fig:F2}B).  The correlated error process rescales the size of all steps in the walk uniformly, and all termination points for a given circuit fall on a line.
The preservation of circuit-structure dependence in the resulting net circuit error leads to a broad distribution of walk lengths that is maintained even when averaging experiments together over various realizations of the random but correlated noise. It is therefore in the distribution of errors over noise-averaged, randomly composed circuits that the signatures of error correlations between gates within a circuit will appear.

Our qubit is realized in the hyperfine $^2\mathrm{S}_{1/2}$ ground state manifold of a laser cooled $^{171}{\textrm{Yb}}^{+}$ ion, with the computational basis states given by \mbox{$\ket{0} \equiv \ket{F=0, \, m_F = 0}$} and \mbox{$\ket{1} \equiv \ket{F= 1, \, m_ F = 0}$}. Transitions between qubit states are driven by a microwave field near $12.6\,$GHz with Rabi frequency $\Omega = 22.5\,$kHz. Engineered noise is added during driven rotations~\cite{SoarePRA2014} via a controlled detuning $\Delta$ of the microwaves from resonance, creating a $\sigz$ off-resonance error with fractional detuning $\delta = {\Delta}/{\Omega}$ simulating the effect of a time-varying magnetic field.  We initially treat two extremal cases: maximally correlated quasi-static noise, $\Mn=J$, which is constant over the entire $J$-gate circuit, and uncorrelated noise, $\Mn \leq 1$, which takes one or more noise values within a single gate and varies randomly between gates.  All noise values are sampled from a Gaussian distribution $\mathcal{N}(0,\sigma^2)$ with zero mean and variance $\sigma^2$. 

We proceed by initializing the qubit in state $\ket{0}$ via optical pumping and implement $k=50$ randomly composed circuits of length $J=100$ that, in the absence of error, each implement the effective target gate $U_{\textrm{T}}=\hat{\mathbb{I}}$. A projective measurement at the end yields the probability of finding the ion in state $\ket{1}$ and the entire experiment is repeated $r=220$ times to reduce quantum projection noise (see \emph{Materials and Methods} for details of the measurement process). We then average the measurement outcomes for each circuit over $n$ different realizations of noise possessing the same engineered correlations. 

Figs.~\ref{Fig:F2}C-E show the distribution of measured noise-averaged probabilities, $\left\langle P(\ket{1})\right\rangle_{n}$, over these random circuits.  The same set of circuits is subject to correlated (gray) or uncorrelated (red) noise sampled from \mbox{$\delta\sim\mathcal{N}(0,\sigma^2=2\times 10^{-3})$}.  Data are represented as histograms for different fixed values of $n$.  As shown in Ref.~\cite{Ball:2016}, despite the difference in noise correlations, the resultant distributions possess approximately the same mean value. Solid lines are theoretical predictions for the distribution over circuits derived from the random walk framework~\cite{Ball:2016, Mavadia:2017}, and show agreement with the data using no free parameters (see \emph{Materials and Methods}). 

The behavior of the distributions under an increasing number of noise averages $n$ is particularly important.  For small $n$ the distributions appear similarly broad, but under further averaging the distribution measured under uncorrelated noise narrows while the variance of the distribution measured using correlated noise remains approximately constant. This is a manifestation of the effect of noise averaging in the presence of the theoretical observations illustrated in Figs.~\ref{Fig:F2}A-B. 

To highlight the effect of noise correlations on the measured averaging behavior, we plot the variance over the distribution of measured 
probabilities \mbox{$\mathbb{V}_\mathrm{k}^{(n)}  \equiv  \mathbb{V}_\mathrm{k}\left[\left\langle P(\ket{1})\right\rangle_{n}\right]$} as a function of the number of noise-averages (Fig.~\ref{Fig:F2}F).  Potential unintended systematic bias in the experimental data's scaling with $n$ is mitigated by random reordering of the experimental outcomes prior to cumulative averaging, producing a spread of individual averaging trajectories. For $\Mn=J$ the resulting trajectories are initially broadly distributed and fluctuate before converging with $n$ to a fixed, analytically calculable variance. By contrast, in the case of uncorrelated noise with $\Mn\leq 1$, all trajectories show an approximate reduction in \mbox{$\mathbb{V}_\mathrm{k}^{(n)} \propto 1/n$}, commensurate with a continued narrowing of the distribution of outcomes over different circuits under averaging (Fig.~\ref{Fig:F2}C-E). 

Solid lines capturing key scaling behaviors observed in both data sets of Fig.~\ref{Fig:F2}F are derived from theoretical predictions. No free parameters are used for the correlated-noise data, while for the uncorrelated data a single overall scaling parameter is fit. Numerical evidence and analytical considerations attribute the initial offset at low $n$ in the correlated error case to contributions from higher order terms (see \emph{Materials and Methods}).
 
The saturation value of {$\mathbb{V}_\mathrm{k}^{(n)}$ achieved in the presence of uncorrelated noise depends on the noise bandwidth and sequence length $J$; we have verified it is not due to fundamental measurement limits in our system or quantum projection noise.  Full details are presented in the \emph{Materials and Methods}. The different scaling behaviors observed under the application of noise processes with different correlation characteristics are reminiscent of the manifestation of noise correlations in other physical quantities, e.g. the Allan variance used in precision frequency metrology~\cite{Allan:1966, Rutman1978}.  Thus we see that this measurement routine provides a clear signature of the presence of temporal correlations between errors resulting from noisy operators applied sequentially to a qubit. 

In an attempt to minimize the propagation of experimental noise correlations to error correlations, we investigate the effect of replacing primitive operations with dynamically corrected gates (DCGs).   We first examine the effect at the operator level using the error vector $\vect{\textbf{a}}_j$ defined above. In the limit of classical Gaussian dephasing noise, described in the Fourier domain as the spectrum $\beta_{z}(\omega)$, the leading order Magnus term ($\alpha=1$) may be written as \mbox{$[\vect{\textbf{a}}_j]_{1}=-i\int \frac{d\omega}{(2\pi)}G^{(1)}_{z}(\omega, T_{j})\beta_{z}(\omega)$}. Here, $G^{(1)}_{z}(\omega, T_{j})$ is an analytically calculable filter transfer function that describes the spectral characteristics of a gate active for duration $T_j$~\cite{ViolaFFF}.  The \emph{effective} error spectrum experienced by the gate may therefore be represented by the spectral overlap of the filter transfer function and noise, written as \mbox{$G_{z}(\omega, T_{j})\times \beta_{z}(\omega)\to E(\omega, T_{j})$}. Figure~\ref{Fig:F3}A demonstrates this mapping schematically for an example spectrum and a ``primitive'' $X_{\pi}$ rotation, where the correlations in the noise are directly transferred to the correlations of the effective error spectrum~\cite{GreenPRL2012} (c.f. Fig.~\ref{Fig:F1}C).

Replacement of a primitive gate with a logically equivalent DCG at the virtual layer modifies the effective error spectrum for each operator~\cite{GreenPRL2012, Kabytayev2014, SoareNatPhys2014, ViolaFFF}.  In Fig.~\ref{Fig:F3}B we illustrate this effect for the CORPSE (Compensation for Off-Resonance with a Pulse SEquence~\cite{True}) protocol, leading to a suppression of low frequency noise that whitens the effective error spectrum relative to $\beta_{z}(\omega)$.  In the current context, this whitening suggests that DCGs should suppress measurement signatures of error correlations between sequentially applied virtual gates.

\begin{figure}[tp!]
\label{fig:3}
\centering
\includegraphics[width=8.3cm]{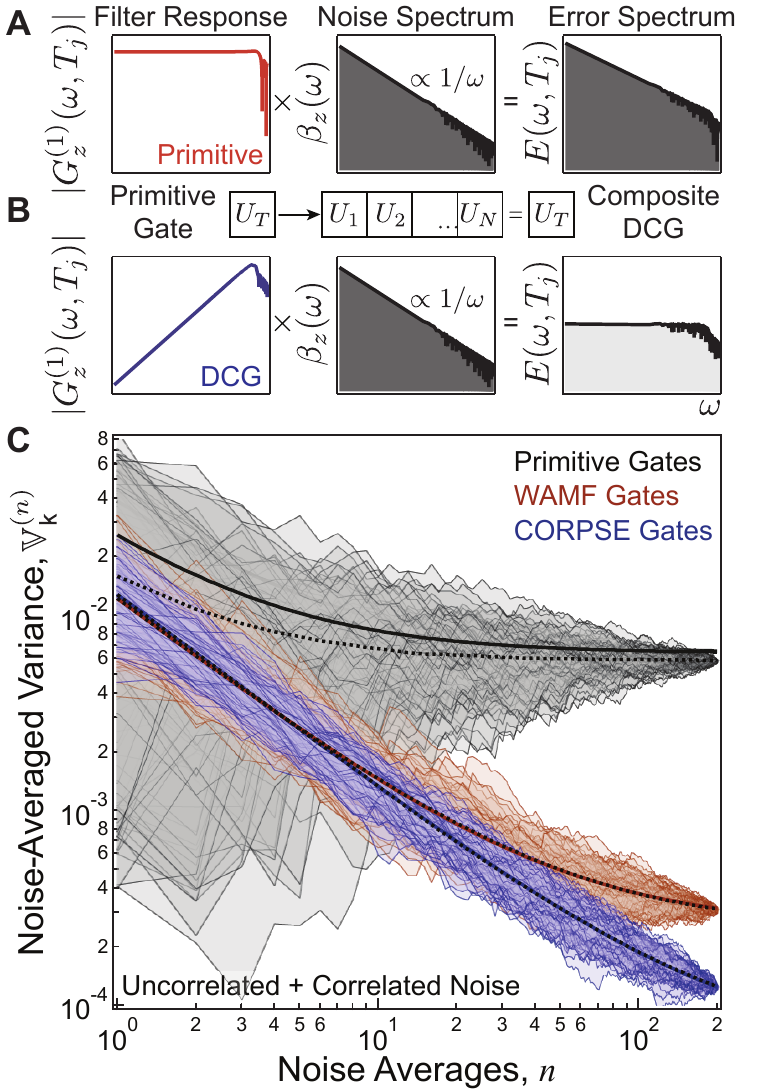}
\centering
\caption{\textbf{Suppression of error correlations using dynamically corrected gates.} (A) The first order, generalized filter transfer function for dephasing noise of a primitive operation, $G^{(1)}_{z}(\omega, T_{j})$, and the noise spectrum (here $\beta_{z}(\omega)\propto1/\omega$) combine to produce an effective error spectrum $E(\omega, T_{j})$ for a single gate.   
(B) The modified filter functions for first order DCGs scale as $\omega$ at low frequencies, which results in a ``whitening'' of $E(\omega, T_{j})$. (C) Variance scaling with $n$ for two different DCG constructions, WAMF (orange) and CORPSE (blue), as well as primitive gates (gray) all subjected to noise with both correlated ($\Mn = J$,  $\delta_L \sim \mathcal{N}(0,\sigLSq=2\times10^{-3}$)) and uncorrelated ($\Mn \leq 1$, $\delta_S \sim \mathcal{N}(0,\sigSSq=5\times10^{-4})$) components.  
Dotted lines are means of 1000 trajectories randomized over noise realizations, and solid lines are theoretical fits to the mean with no free parameters for the primitive data and extracting estimates of $\sigSSq$ and $\sigLSq$ from the DCG data.  The best fit for CORPSE results in an estimate of $\sigLSq = 5.6 \times10^{-6}$ and $\sigSSq = 7.3 \times10^{-3}$ with $R^{2}=0.999989$.  We employ the Akaike information criterion to bound a confidence interval for competing models using different (free) values of $\sigSSq$ and fixed varying $\sigLSq$.   This gives a 5\% relative likelihood bound for $\sigLSq =  (5.6 \protect\substack{+1.9 \\ -2.3} ) \times10^{-6}$.  The Bayesian information criterion exceeds 10 (strong model rejection) for values outside the interval $\sigLSq  = (5.6 \protect\substack{+2.6 \\ -3.2} ) \times10^{-6}$.  See \emph{Materials and Methods} for full details of statistical analysis.
}
\label{Fig:F3}
\end{figure}

We experimentally implement primitive, CORPSE and WAMF (Walsh Amplitude-Modulated Filter~\cite{Ball:2014}) gates, where both DCGs are designed to suppress errors due to detuning noise. Using the same randomly composed circuits as in Fig.~\ref{Fig:F2}, we now apply a mixed noise spectrum simultaneously containing uncorrelated, rapidly varying noise \mbox{($\Mn \leq 1$)} sampled from $\delta_S \sim \mathcal{N}(0,\sigSSq)$, and offsets that are constant over a single circuit giving a strongly correlated component \mbox{($\Mn = J$)} with $\delta_L \sim \mathcal{N}(0,\sigLSq)$.   Introducing DCGs and measuring the averaging behavior of $\mathbb{V}_\mathrm{k}^{(n)}$ thus permits measurement of the impact of modifying the effective error spectrum on the manifestation of error correlations between virtual gates in the circuit.  

Both DCG implementations show initial variance scaling with noise averaging $\mathbb{V}_\mathrm{k}^{(n)}\propto 1/n$ \mbox{(Fig.~\ref{Fig:F3}C)}, reminiscent of the application of an uncorrelated noise process in Fig.~\ref{Fig:F2}F.  The observed saturation at large $n$ for the DCG data combines contributions due to both residual error correlations and the analytically calculable saturation occurring in the presence of purely uncorrelated errors introduced above.  This general behavior is to be contrasted with that observed for the same circuits composed of primitive gates, where the strong correlated noise again gives rise to a variance that converges to a large constant value (gray).

We combine the theoretical predictions for the scaling of $\mathbb{V}_\mathrm{k}^{(n)}$ with $n$ for the primitive gates using an analytic model that incorporates both effective noise contributions -- correlated and uncorrelated -- using no free parameters (see \emph{Materials and Methods}).  The resultant model agrees with the data to within 10\% of the saturation value. We believe residual disagreement comes from the treatment of higher order error covariances in the calculation when considering simultaneous correlated and uncorrelated errors. 

In order to deduce the change in relative correlated and uncorrelated error components after DCG application, we now employ this model to the DCG data, allowing $\sigSSq$ and $\sigLSq$ to vary freely, and extract estimates for the effective $\sigSSq$ and $\sigLSq$ from the best fits. The model is adjusted to account for the increased bandwidth of the uncorrelated noise component relative to the gate length, with the noise now changing $\sim8\times$ in a DCG gate. First, for both DCGs we find an increase in the extracted uncorrelated error component, $\sigSSq$, by a factor of $\sim 14-16$ relative to the applied noise process.  The effective correlated error, $\sigLSq$, however, is reduced by a factor of $370\times$ for CORPSE and $16\times$ using WAMF gates (see \emph{Materials and Methods}).  

The relative performance of these two DCGs observed in our experiments is aligned with their documented strengths, as CORPSE is known to more efficiently cancel purely static detuning errors than WAMF~\cite{Kabytayev2014, Ball:2014}, although improved calibration of the pulse-amplitude values used in WAMF gates is expected to improve the relatively poor correlated-error suppression observed here. Moreover the high-pass filtering nature of both DCGs illustrates why uncorrelated noise processes fluctuating rapidly on the scale of the individual DCGs lead to residual errors that are transmitted through the filter.  These measurements -- in particular the scaling of $\mathbb{V}_\mathrm{k}^{(n)}$ -- are consistent with an interpretation that the action of the noise whitening in the filter-transfer-function framework transforms correlated noise into predominantly uncorrelated residual errors at the operator level  
(see \emph{Materials and Methods} for full details).

\begin{figure}[bt]
\label{fig:4}
\centering
\includegraphics[width=8.3cm]{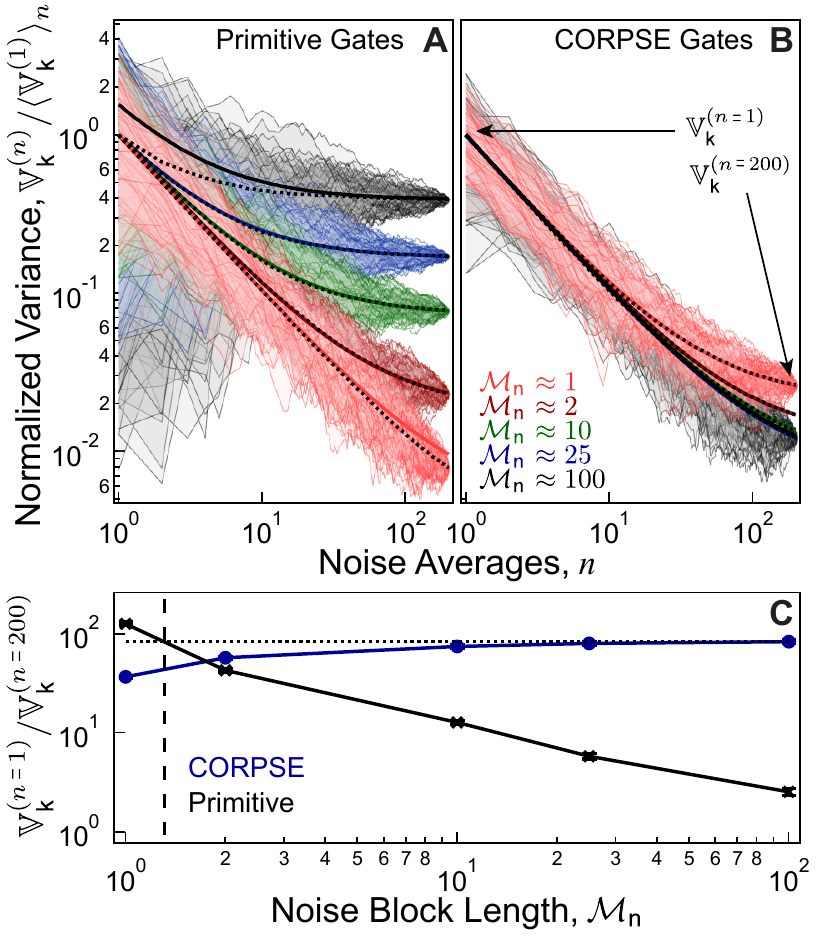}
\caption{\textbf{Suppression of error correlations using DCGs under noise with varying $\Mn$.} (A-B) Variance scaling of $k=20$ circuits with noise averaging for (A) primitive and (B) CORPSE gates. Traces are normalized to the initial mean variance for each applied noise case. Engineered noise is composed of an uncorrelated component ($\Mn \leq 1$) and a block correlated component of length $\Mn$ that is varied from fully correlated ($\Mn = J$) to uncorrelated ($\Mn = 1$) between virtual gates. Solid lines correspond to analytic fits where the correlated and uncorrelated error strengths are allowed to vary.
(C) Ratio of initial to final variance in the upper panels as $\Mn$ is varied for primitive (black) and CORPSE (blue) gates. Dotted line marks the ratio at which CORPSE gates saturate, dashed vertical line indicates value of $\Mn$ where this value crosses the scaling trend for primitive gates. Error bars calculated from the SEM of the 200 initial values of variance and normalized by the fully noise-averaged variance are smaller than point-size.} \label{Fig:F4}
\end{figure}

We conclude by experimentally demonstrating that the reduction in effective error correlation, indeed, resides at the virtual gate layer. Using the same circuits as before and the same correlated and uncorrelated noise strengths, we now vary the length of the correlated noise component at the virtual level, breaking it up into blocks of length $\Mn$.The physical lengths of the noise blocks therefore differs by a factor of $\sim6$ between the primitive and the CORPSE gates (the average increase in gate length for the 24 Clifford operations). In the case of primitive gates, the signature exhibited by the variance scaling under noise averaging in Fig.~\ref{Fig:F4}A gradually changes from indicating uncorrelated errors ($1/n$-like scaling) to correlated errors (saturation at high variance) as the block length is increased. By contrast, the CORPSE gates in Fig.~\ref{Fig:F4}B retain their overall $1/n$-like scaling behavior for all $\Mn$, demonstrating that residual uncorrelated errors remain dominant. The traces here have been normalized to the initial mean variance for each engineered noise case to highlight the change in the relative correlated and uncorrelated error components, rather than the net error strength. 

As a witness of this behavior, Fig.~\ref{Fig:F4}C shows the ratio of the initial mean variance, $\mathbb{V}_\mathrm{k}^{(n=1)}$, to the final, fully noise-averaged variance, $\mathbb{V}_\mathrm{k}^{(n=200)}$. This ratio scales approximately inversely with $\Mn$ for primitive gates but remains approximately constant for CORPSE gates.  Extrapolation of the saturation value of this ratio for CORPSE back towards small $\Mn$ reveals a crossover with the primitive data that lies between $\Mn =1-2$, confirming CORPSE's ability to suppress error correlations between virtual gates.

The results we have presented above suggest that the path to the practical implementation of QEC may be facilitated by transforming common laboratory noise sources exhibiting slow drifts to effective error processes with dramatically reduced correlations at the virtual layer using DCGs.  We believe this is important as the pursuit of functional quantum computers - even at the mesoscale - will clearly require major advances in the control and suppression of errors with gate counts for even moderate problems requiring only $\sim200$ qubits quickly exceed $10^{10}$~\cite{Reiher:2017}.  Future work will involve exploration of the efficacy of a range of DCG constructs in suppressing temporal and spatial correlations under complex classical and quantum error models.  Combined with the observation that certain DCGs can mitigate spatial crosstalk in multiqubit systems~\cite{Brown_addressing}, we believe that our demonstration of the suppression of temporal error correlations within quantum circuits solidifies the central importance of dynamic error suppression techniques at the virtual level for practical quantum computing.  

\emph{Acknowledgements}: The authors acknowledge S. Mavadia for assistance with data collection and simulations, and discussions with C. Ferrie, and C. Granade on data analysis. Work partially supported by the ARC Centre of Excellence for Engineered Quantum Systems CE110001013, the Intelligence Advanced Research Projects Activity (IARPA) through the US Army Research Office, and a private grant from H. \& A. Harley.
\newpage

\clearpage
\begin{widetext}
\appendix
\begin{center}
\LARGE{Supplementary Materials}
\end{center}
\section{Experimental Setup} 
%

Our qubit is encoded in the $^2\mathrm{S}_{1/2}$ hyperfine ground states of a single laser-cooled ${}^{171}{\textrm{Yb}}^{+}$ ion confined in a linear Paul trap, with the computational basis states defined as $\ket{0} \equiv \ket{\textrm{F}=0, \textrm{m}_\textrm{F} = 0}$ and $\ket{1} \equiv  \ket{\textrm{F}=1, \textrm{m}_\textrm{F} = 0}$.
Laser cooling, state initialization to $\ket{0}$ and detection are performed using a laser at 369.4\,nm, which is coupling the $\ket{{}^{2}{\textrm{S}}_{1/2}, \textrm{F}=1}$ ground state to the first excited state $\ket{^2\mathrm{P}_{1/2}, \textrm{F}=0}$.

As the ion selectively fluoresces when it is projected to the upper, bright qubit state $\ket{1}$, we are able to distinguish between the two basis states by counting the number of emitted photons during the detection period.
Further details about the state detection protocol, including a Bayesian inference procedure used to determine the state from both the number of counted photons and their arrival times, can be found in the \emph{Supplementary Materials} of reference \cite{Mavadia:2017}. 

Single-qubit rotations are driven via a microwave field near 12.6~GHz generated by a Keysight E8267D Vector Signal Generator (VSG). 
Using the built-in $IQ$ modulation, we can freely adjust the phase of the microwave signal to implement rotations around arbitrary equatorial axes on the Bloch sphere. Rotations about the $z$-axis are implemented as instantaneous, pre-calculated $IQ$ frame shifts and consequently are not susceptible to engineered detuning or amplitude errors. 

Quantum circuits of multiple Clifford operations are preloaded into the VSG and selectively compiled prior to the recording of each data set. The corresponding microwave pulses are switched using an inbuilt VSG protocol, RF blanking, which serves to minimize microwave leakage between operations and at the end of a circuit. In addition, the technique suppresses ringing in the pulse amplitude of the microwaves at the beginning and conclusion of an operation, which is caused by updates of the $IQ$ values defining both phase and amplitude.

\section{Measurement Procedure and Engineered Noise Correlations}
The experiments in this manuscript are performed using $k=50$ circuits each comprising $J=100$ operations. Here, the first $J-1=99$ are randomly composed Clifford operations $\mathcal{\hat{C}}_j$ and the final operation $\mathcal{\hat{C}}_J =  (\prod_{j=1}^{J-1}\mathcal{\hat{C}}_j)^{\dagger}$ is selected such that the circuit implements the identity $\Identity$ in the absence of error. 
A full list of the Clifford operations and their physical implementation can be found in the \emph{Supplementary Materials} of reference \cite{Ball:2016}.

Each circuit is executed in the presence of engineered detuning noise characterized by a temporal correlation length $\Mn$. In particular, three cases are implemented: (1) fully correlated across the circuit ($\Mn = J$), (2) fully uncorrelated between sequential gates, with noise values stochastically varying in blocks commensurate with the length of a physical $\pi/2$ rotation ($\Mn\leq1$), and (3) a combination of both correlated and uncorrelated noise components ($\Mn^\mathrm{cor} \geq \Mn^\mathrm{uncorr}\leq1$). 

This type of noise process is informed by the realistic situation of a time varying magnetic field, which changes the qubit energy splitting, creating a detuning from the driving field. In particular, fluctuations at 50~Hz (resp. 60~Hz) can be commonly observed due to the presence of AC mains connections. Other strongly correlated slow frequency drifts are often related to changes associated with the ambient temperature of electrical equipment and duty cycle changes during operations, while fast fluctuations are usually caused by electrical noise in components that is insufficiently filtered or intrinsic to the qubit environment (such as TLS noise in superconducting qubits or anomalous heating in ion traps). In general, detuning noise will therefore both have a correlated (slow) component and a \enquote{fast} largely uncorrelated component.

For each instance of engineered noise, $n=200$ noise realizations were sampled from a normal distribution $\delta\sim\mathcal{N}(0,\sigma^2)$ with rms $\sigma$. Here, $\delta = (\Delta/\Omega)$ is a  fractional detuning expressed by the ratio of the frequency detuning $\Delta$ from the qubit transition frequency near 12.6~GHz normalized by the Rabi frequency $\Omega$ (coupling strength) of a driven rotation. Every combination of circuit and noise realization was repeated $r=220$ times to reduce the impact of quantum projection noise.

%
\section{Dynamically Corrected Gates}
\label{sec:DCG}
%

Dynamically corrected gates (DCGs) are implemented by replacing \enquote{primitive} physical rotations with composite sequences comprised of multiple physical rotations \cite{Kabytayev2014}. In particular, we are investigating the \enquote{Compensation for Off-Resonance with a Pulse SEquence} (CORPSE) \cite{Cummins:2000,True} and \enquote{Walsh amplitude modulated filter} (WAMF) \cite{Ball:2014} approaches that abstract target rotations away from the underlying physical operations to a virtual gate level. 

In both cases, target $\theta_t = \pi$ and $\theta_t = \pi/2$ gates are constructed as 3 segment pulses with the segments' rotation angles $\theta_i$, Rabi frequencies $\Omega_i$ relative to some maximum frequency $\Omega$, and phase angles $\phi_i$ as indicated in the table below.

\begin{table}[H]
\begin{centering}
\renewcommand{\arraystretch}{2}
      \begin{tabular}{|c||c|c|c|}
  \hline
 	DCG type& ($\theta_1, \Omega_1, \phi_1$) &($\theta_2, \Omega_2, \phi_2$) & ($\theta_3, \Omega_3, \phi_3$) \\ \hline\hline
CORPSE	& ($2\pi + \theta_t/2 - k , \Omega, 0 $)	& 	($2\pi - 2k , \Omega, \pi $)  & ($\theta_t/2 - k , \Omega, 0 $)	 \\ \hline
WAMF 	& ($\tfrac{X_0 + X_3}{4} , \Omega, 0 $)	& 	($\tfrac{X_0 - X_3}{2} , \tfrac{X_0 - X_3}{X_0 + X_3} \Omega, 0 $)  & ($\tfrac{X_0 + X_3}{4} , \Omega, 0 $)	 \\ \hline
    \end{tabular}\caption{Required parameters to construct a target $\sigx$ CORPSE and WAMF rotation with target angle $\theta_t$. An additional $\pi/2$ shift in $\phi$ is required for $\sigy$ rotations. Here, $k = \arcsin{(\tfrac{\sin{(\theta_t/2)}}{2})}$ and for WAMF DCGs, the target rotations $\theta_t = (\tfrac{\pi}{4}, \tfrac{\pi}{2}, \pi)$ have $X_0 = (2\tfrac{1}{4}, 2\tfrac{1}{2}, 3)\pi$ and $X_3 = (0.36, 0.64, 1)\pi$ determined explicitly.}\label{SuppTable:DCGs}
\end{centering}
\end{table}

A schematic of the gates for a target rotation $\theta_t = \pi$ about the $x$-axis is shown in Fig. \ref{SuppFig:DCGs}.
\begin{figure}[h]
\centering
\includegraphics[scale = 1]{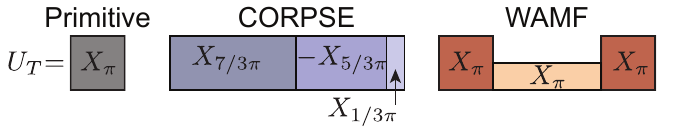}
\caption{Construction of a CORPSE and WAMF DCG for target rotation $\theta_t = \pi$ about the $x$-axis.
}
\label{SuppFig:DCGs}
\end{figure}

To ensure that the error suppressing aspects of the DCGs are maintained for all Clifford gates, we implement their identities $\Identity$ by concatenating an $X_\pi$ rotation and its inverse $-X_\pi$ in the case of CORPSE and WAMF. While this again results in a net zero rotation, effectively identical to the simple wait time of \enquote{primitive} gates, it makes the identity operation first-order insensitive to detuning errors during its operation.

%
\section{Linking Noise to Error Using the Error Vector}
%

Any noisy operation, $\tilde{U}_j$, can be decomposed into an ideal component, $U_j$ and an error component, $\tilde{U}_{\varepsilon,j}$, such that \mbox{$\tilde{U}_j = \tilde{U}_{\varepsilon,j}U_j$}. The error operator is expressed as $\tilde{U}_{\varepsilon,j} = \textrm{exp}\{ \sum_{\alpha=1}^\infty [\errorVect_j]_\alpha \cdot \vec{\sigma}  \}$, where $\alpha$ denotes the order of the so-called Magnus expansion, $\vec{\sigma}$ is the vector of Pauli matrices associated with the operation, and $\errorVect_j$ is the error vector characterizing the strength and nature (affected quadrature) of the error. Correlations in the error process manifest in its gate-by-gate evolution throughout a quantum circuit.

At the level of physical gate rotations (\enquote{primitive} gates), one can observe a direct translation between correlations in an applied detuning error process during a quantum circuit and correlations arising in the magnitude of the error vector for each of the gates in the circuit. 
Supported by the filter-transfer-function framework \cite{ViolaFFF}, this suggests that primitive gates map noise correlations directly to error correlations. 

In the main text, we show this mapping by calculating the autocorrelation function of the first-order error vector magnitude for a single randomly composed circuit under one noise instance with varying block correlation length, $\Mn$.
In Figure~\ref{SuppFig:ErrorVect} below, this mapping is seen to be persistent, if not strengthened, even under averaging error vectors over randomly composed circuits and noise realizations.

\begin{figure}[h!]
\centering
\includegraphics[scale = 1]{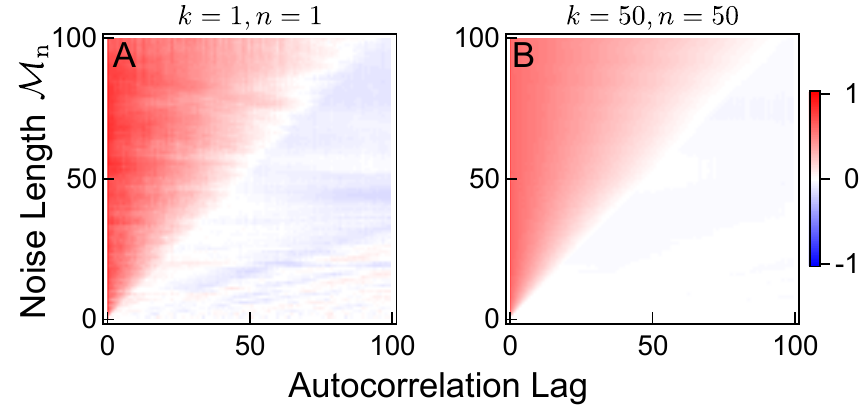}
\caption{\textbf{Mapping correlations in a noise process to correlations in circuit errors}
Autocorrelation function (ACF) of the first-order error vector magnitude, $\vert \errorVect \vert$, for the first 100 gates of a circuit comprising $J=1000$ randomly composed primitive Clifford operations under a detuning noise process with fractional detuning $\delta \sim \mathcal{N}(0,\sigma^2)$, which changes value every noise block length $\Mn$. (A) ACF calculated for a single $J=1000$ gate circuit under one noise realization, and (B) ACF averaged over $k=50$ circuits and $n=50$ noise realizations to illustrate the persistence the mapping from noise correlations $\Mn$ to error correlations $\Me$ for primitive gates.} 
\label{SuppFig:ErrorVect}
\end{figure}
%

%
\section{Modifications to Original Random Walk Model for Randomized Benchmarking}
%


The theoretical model underlying this work was initially presented by Ball et al. in reference \cite{Ball:2016}, wherein the error process studied described an instantaneous phase error, $e^{i\delta\sigma_z}$, occuring after each Clifford operation in a randomly composed quantum circuit, such as those used in Randomized Benchmarking (RB). In this context, the dephasing magnitude was sampled from a zero-mean Gaussian distribution with rms $\sigma$, $\delta\sim\mathcal{N}(0,\sigma^2)$. 

The key finding of Ball et al. was that, to first order, it is possible to map between the errors occurring throughout the Clifford operation circuit and a walk in 3D Pauli error space, with the net walk length relating to the circuit fidelity. 
It was found that this error process results in noise-averaged circuit fidelities that are Gamma distributed, $\noiseAve{\mathcal{F} } \sim \Gamma(\alpha,\beta)$. The shape and scale parameters, $\alpha$ and $\beta$ respectively, can be calculated from first principles using the strength of the error process $\sigma^2$, the circuit length $J$, and the number of noise averages $n$. These parameters describe how the mean and width of the distribution change with noise averaging. The distributions for errors that are constant across a circuit (correlated) and errors that change randomly between sequential gates (uncorrelated) are respectively given by 
\begin{align}
1 - \langle \mathcal{F} \rangle_{n,\textrm{correlated}} &\sim \Gamma(\alpha = \tfrac{3}{2}, \beta = \tfrac{2}{3}J\sigma^2) \\
1 - \langle \mathcal{F} \rangle_{n,\textrm{uncorrelated}} &\sim \Gamma(\alpha = \tfrac{3n}{2}, \beta = \tfrac{2}{3n}J\sigma^2)
\end{align}
where the distribution variance and expectation are given by $\mathbb{E} = \alpha\beta, \mathbb{V} = \alpha\beta^2$. Consequently, we see there is narrowing of the distribution with noise averaging soley for uncorrelated errors.
In the following, we present a revised version of this theoretical model linking it to the experiments performed in the present manuscript. 

%
\subsection{Model Revision for Survival Probability Measurement}
%

When applying the above theory to experimental results, it is necessary to consider how the measurement protocol differs from the analytic model. The original theory was based around the fidelity of a noisy circuit operation, with net operator $\RBseqNoisy$, compared to the ideal circuit, $\RBseqClean = \Identity$,
\begin{align}
\mathcal{F} &= \tfrac{1}{4} \vert \textrm{Tr} (\RBseqClean^\dagger \RBseqNoisy) \vert^2 \nonumber \\
&= \tfrac{1}{4} \vert \textrm{Tr} (\RBseqNoisy) \vert^2.
\end{align}
In our experiment we measure the probability $P(\ket{1})$ of a qubit initially prepared in $\ket{0}$ \emph{not} to return to $\ket{0}$ but end up in state $\ket{1}$. As this is a projective measurement onto the $z$-axis of the Bloch sphere, it is insensitive to rotations about the $z$-axis, \emph{i.e.} it is phase-invariant. Consequently, we are insensitive to the component of the Pauli space walk in the $\sigz$-direction. Indeed, the projective measurements actually probe a 2D projection of the walk onto the $\sigx,\sigy$-plane, and the Gamma distribution shape and scale parameters become
\begin{align}
\langle P(\ket{1}) \rangle_{n,\textrm{correlated}} &\sim \Gamma(\alpha = 1, \beta = \tfrac{2}{3}J\sigma^2) \label{eq:2DCorGamma}  \\
\langle P(\ket{1})) \rangle_{n,\textrm{uncorrelated}} &\sim \Gamma(\alpha = n, \beta = \tfrac{2}{3n}J\sigma^2) \label{eq:2DUncorGamma}.
\end{align}
Further details can be found in the \emph{Supplementary Materials} of \cite{Mavadia:2017}. 

%
\subsection{Model Revision for Concurrent Detuning Error Processes}
%
A second alteration to the model emerges from our method of noise engineering. The results presented here study a time varying or constant frequency detuning during the circuit's execution. Unlike in the original model, this induces multi-axis errors throughout the individual Clifford operations, not just between them. 

Due to the gates spanning different lengths for $\pi$ and $\pi/2$ rotations, they accumulate different amounts of phase from the detuning. As such, the analytic model must now consider gate-dependent errors. Such errors violate the original assumptions of randomized benchmarking as has recently been highlighted in reference \cite{Wallman:2017}.

The adaptation to the theory for noisy Clifford gates is initially presented for two noise processes: (1) detunings that are constant across individual gates but vary randomly between gates, or (2) constant detunings across the entire circuit, giving maximal temporal correlation in the noise. 


Starting with the standard randomized benchmarking procedure we compile a circuit of randomly composed single qubit Clifford operations
$\prod_{j-1}^J \hat{C}_j = \mathbb{I}$
such that in the absence of error the final state will be the same as the prepared state.
\noindent The effectively implemented gates, $\tilde{C}_j$, differ from the ideal gates $\hat{C}_j$ by an error map $\Lambda_j$ that satisfies $\tilde{C}_j=\Lambda_j \hat{C}_j$. Then, the circuit is given by
\begin{equation}
\prod_{j=1}^J \Lambda_j\hat{C}_j=\tilde{S}.
\end{equation}

\noindent The single qubit Clifford gates are made up of rotations about the Bloch sphere
\begin{equation}
\hat{R}_{\hat{W}}(\theta)=e^{-i\tfrac{\theta}{2}\hat{W}},
\label{eq:idealrot}
\end{equation}
where $\hat{W}\in \{\mathbb{I},\sigx,\sigy,\sigz\}$ and $\theta=\pi, \pm\tfrac{\pi}{2}$. 

\noindent The implemented rotations, with the engineered error, are
\begin{equation}
\tilde{R}_{\hat{W}}(\theta,\delta)=e^{-i\left(\tfrac{\theta}{2}\hat{W}+\tfrac{\lvert\theta\rvert}{2}\delta \sigz\right)},
\label{eq:tilderot}
\end{equation}
except for $\sigz$ rotations which are implemented as a passive frame change, hence error free. Using the standard definition of single qubit Clifford gates \cite{Ball:2016}, there is only one non $\sigz$ rotation, hence one error map, per gate.

\noindent We calculate the error map for the different rotations 
\begin{subequations}
\label{eq:decomposed_all}
\begin{align}
\Lambda^{(\mathbb{I})}(\pi,\delta)&=\left(1-\tfrac{\pi^2 \delta^2}{8}\right)\mathbb{I}-i\tfrac{\pi\delta}{2}\sigz+ \mathcal{O}(\delta^3),\\
\Lambda^{(\X)}(\pi,\delta)&= \left(1-\tfrac{\delta ^2}{2}\right)\mathbb{I} -\tfrac{1}{4} i \pi  \delta ^2 \sigx +i \delta \sigy+ \mathcal{O}(\delta^3),\\
\Lambda^{(\X)}(\pm\tfrac{\pi}{2},\delta)&= \left(1-\tfrac{\delta ^2}{4}\right)\mathbb{I}\pm\tfrac{2-\pi}{8} i  \delta ^2\sigx\pm\tfrac{i \delta }{2}\sigy-\tfrac{i \delta }{2}\sigz+ \mathcal{O}(\delta^3),\\
\Lambda^{(\Y)}(\pi,\delta)&=\left(1-\tfrac{\delta ^2}{2}\right)\mathbb{I}-i \delta \sigx-\tfrac{1}{4} i \pi  \delta ^2\sigy+ \mathcal{O}(\delta^3),\\
\Lambda^{(\Y)}(\pm\tfrac{\pi}{2},\delta)&=\left(1-\tfrac{\delta ^2}{4}\right)\mathbb{I}\mp\tfrac{i \delta }{2}\sigx\pm\tfrac{2-\pi}{8} i \delta ^2\sigy-\tfrac{i \delta }{2}\sigz+ \mathcal{O}(\delta^3),
\end{align}
\end{subequations} 
which can be written in the general form
 \begin{equation}
 \Lambda_j=\mathbb{I}+i \delta_j \bm{\nu}_j \cdot \bm{\sigma}+ \delta^2(i \bm{\eta}_j\cdot \bm{\sigma} - a_j \mathbb{I}) + \mathcal{O}(\delta^3),
 \label{eq:genmap}
 \end{equation}
where $\bm{\sigma}$ is a vector of Pauli matrices and the vectors $\bm{\nu}$, $\bm{\eta}$ and $a_j$ depend on which error map from \eqn{eq:decomposed_all} is used. 

\noindent The survival probability averaged over $n$ noise instances is calculated using
\begin{equation}
1-\noiseAve{ P(\ket{1}) }=\noiseAve{\lvert\bra{0}\tilde{S}\ket{0}\rvert^2}.
\end{equation}
We use the method from \cite{Ball:2016} to approximate the circuit. Each error map can be translated to a step in Pauli space away from the ideal state, with the total, noise-averaged, random walk given by
\begin{equation}
\noiseAve{\bm{R}}=\tfrac{1}{n}\sum_{k=1}^n\sum_{j=1}^J \delta_{jk} \bm{r}_j.
\end{equation}
Here $\bm{r}_j$ is from the product of the preceeding and succeeding ideal gates modifying $\bm{\nu}_j$ as so $\hat{C}_{1...j-1}\bm{\nu}_j\cdot\bm{\sigma}\hat{C}_{1...j-1}^\dagger=\bm{r}_j\cdot\bm{\sigma}$. 

\noindent It can be shown \cite{Harris:2018} that the survival probability is given by
\begin{equation}
1 - \noiseAve{ P(\ket{1}) }=1-\left(\noiseAve{ \lvert\bm{R}\rvert^2} - \noiseAve{\lvert R_z\rvert^2}+\mathcal{O}(\delta^3)\right),
\end{equation}
where $R_z$ is the walk along the $\sigz$-axis in Pauli space.

\noindent We calculate the characteristics of the survival probability from the statistics of the walk, weighting the contribution of each gate type by the gate-dependent step \mbox{$1\hat{n}_1, (\tfrac{1}{2}\hat{n}_1 +  \tfrac{1}{2}\hat{n}_2), \tfrac{\pi}{2}\hat{n}_1$} for $\pi, \tfrac{\pi}{2}, \Identity$ gates respectively with $\hat{n}_1,\hat{n}_2 \in \{\sigx,\sigy,\sigz\}$, and get the expectation value
\begin{equation}
	\E{\noiseAve{ P(\ket{1}) }}\approx J \sigma^2 \tfrac{2}{3}\left(\tfrac{1}{2}+\tfrac{\pi^2}{96}\right). \label{eq:meanstdc}
	\end{equation}
For the noise-averaged variance we need to take correlations into account due to the gate dependant nature of the error maps. The gate dependance gives us a random number of steps along a random axis, which leads to correlations even after averaging over different step lengths.

\noindent When we have uncorrelated errors, we calculate the variance to be
\begin{align}
\Var{\noiseAve{ P(\ket{1}) }}\approx&\tfrac{J^2 \sigma^4}{n}\left(\tfrac{4}{9}\left(\tfrac{1}{2}+\tfrac{\pi^2}{96}\right)^2+\tfrac{1}{J}\left(3\left(\tfrac{7}{36}+\tfrac{\pi^4}{576}\right)\nonumber \right.\right.\\
&\left.\left.-\tfrac{8}{9}\left(\tfrac{1}{2}+\tfrac{\pi^2}{96}\right)^2\right)+\tfrac{(n-1)}{J}\left(\tfrac{7}{36}+\tfrac{\pi^4}{576}\right.\right.\nonumber \\
&\left.\left.-\tfrac{4}{9}\left(\tfrac{1}{2}+\tfrac{\pi^2}{96}\right)^2\right)\right), \label{eq:varstac}
\end{align}
noting that in the limit $n\rightarrow\infty$, the variance scaling saturates at a constant $\propto \tfrac{1}{J}$.

\noindent For correlated errors we get
\begin{align}
\Var{\noiseAve{ P(\ket{1}) }}\approx&\tfrac{J^2 \sigma^4}{n}\left(\tfrac{12}{9}\left(\tfrac{1}{2}+\tfrac{\pi^2}{96}\right)^2+\tfrac{1}{J}\left(3\left(\tfrac{7}{36}+\tfrac{\pi^4}{576}\right)\nonumber \right.\right.\\
&\left.\left.-\tfrac{8}{3}\left(\tfrac{1}{2}+\tfrac{\pi^2}{96}\right)^2\right)+(n-1)\left(\tfrac{4}{9}\left(\tfrac{1}{2}+\tfrac{\pi^2}{96}\right)^2+\nonumber \right.\right.\\
&\left.\left.\tfrac{1}{J}\left(\tfrac{7}{36}+\tfrac{\pi^4}{576}-\tfrac{8}{9}\left(\tfrac{1}{2}+\tfrac{\pi^2}{96}\right)^2\right)\right)\right), \label{eq:varstdc}
\end{align}
again tending towards a constant which, however, now occurs at a significantly smaller number of noise averages than seen previously.


Using the revised model, the noise-averaged survival probability distributions under correlated noise remain Gamma distributed, with an updated scale parameter. 
While this is yet to be shown explicitly for the uncorrelated case, we can approximate its behavior in the limit $n<J$ by modifying the distribution in \eqref{eq:2DUncorGamma}, yielding 
\begin{align}
\langle P(\ket{1}) \rangle_{n,\textrm{correlated}} &\sim \Gamma(\alpha = 1, \beta = \tfrac{2}{3}J\sigma^2 (\tfrac{1}{2}+\tfrac{\pi^2}{96}) ) \label{eq:2DCorConGamma}  \\
\langle P(\ket{1})) \rangle_{n,\textrm{uncorrelated}} &\sim \Gamma(\alpha = n, \beta = \tfrac{2}{3n}J\sigma^2  (\tfrac{1}{2}+\tfrac{\pi^2}{96}) ) \label{eq:2DUncorOneConGamma}.
\end{align}
The renormalized Gamma distributions for correlated error processes shown by solid gray lines in the main text Figs. 2C-E were calculated from first principles using \eqref{eq:2DCorConGamma} with no free parameters. The distributions for the uncorrelated error process in red were calculated from an altered version of \eqref{eq:2DUncorOneConGamma}, which was modified for higher bandwidth noise as explored below.


We note that there is a deviation between the theory and the experiment for correlated noise at early values of $n$, as shown in main text Fig.~2F. 
 However, crucially, even when we update the model with additional fit factors to account for this early $n$ scaling in main text Fig. 2F, the extracted values of $\sigLSq,\sigSSq$ for both DCGs in main text Fig. 3C are completely unaltered within the confidence bounds calculated in Section~\ref{sec:statistical_tests} below.

%
\subsection{Higher Bandwidth Uncorrelated Noise Processes}
%

The engineered uncorrelated noise process in this work have a higher bandwidth than that treated in the error model above. The noise was engineered to change stochastically every primitive $\pi/2$ time, leading to noise that took two values in primitive $\pi$ and $\Identity$ gates, and one value in primitive $\pi/2$ gates. In CORPSE DCGs, the noise took approximately 8 values in both $\pi$ and $\pi/2$ due to the increased length of the virtual gates and 16 values in (virtual) $\Identity$ gates, which were constructed as a composite sequence of $X_\pi$ followed by $-X_\pi$. As this work attempted to quantitatively extract error strengths from the variance scaling trends, it was necessary to update the model to account for this increased noise bandwidth relative to the gate length.

\noindent We recalculate the error map for the $\pi$ gate using two sequential $\pi/2$ gates, each with a different fractional detuning $\delta_1, \delta_2\sim \mathcal{N}(0,\sigma^2)$,
\begin{align}
\Lambda^{(\X)}(\pi,\delta_{1,2}) &= (\mathbb{I} + \tfrac{i (\delta_1 + \delta_2)}{2}\sigy - \tfrac{i (\delta_1 - \delta_2)}{2}\sigz+ \mathcal{O}(\delta^2)) \nonumber   \\
&\equiv (\mathbb{I} + \tfrac{i \delta}{\sqrt{2}}\sigy - \tfrac{i \delta}{\sqrt{2}}\sigz+ \mathcal{O}(\delta^2)).
\end{align}
where $\delta \sim \mathcal{N}(0,\sigma^2)$. This equivalence occurs because $\delta_1,\delta_2$ are independent samples from a Gaussian distribution, meaning their combination is also Gaussian distributed,
\begin{align}
A\delta_1 \pm B\delta_2 &\sim \mathcal{N}(0,A^2\sigma^2) + \mathcal{N}(0,(\pm B)^2 \sigma^2) \nonumber  \\
&= \mathcal{N}(0,(A^2 + B^2)\sigma^2).
\label{eq:gaussSum}
\end{align}
Hence, $A=B=1$ and we can alternatively express this as
\begin{align}
\delta_1 \pm \delta_2 &\equiv \sqrt{2}\delta \nonumber  \\
&\sim \sqrt{2}\mathcal{N}(0,\sigma^2) \nonumber  \\
&= \mathcal{N}(0,2\sigma^2)
\end{align}
where $\delta \sim \mathcal{N}(0,\sigma^2)$. Therefore, we simply adjust the step length for a $\pi$ gate from 1 along a single axis to $\tfrac{1}{\sqrt{2}}$ along two axes, or 1 between two axes, when the noise bandwidth is increased to taking two values per gate.
Similarly, the $\Identity$ gate error map can be rewritten as
\begin{align}
\Lambda^{(\mathbb{I})}(\pi,\delta)&=\mathbb{I}-i\tfrac{\pi(\delta_1+\delta_2)}{4}\sigz+ \mathcal{O}(\delta^2) \nonumber  \\
&\equiv    \mathbb{I}-i\tfrac{\pi\delta}{2\sqrt{2}}\sigz+ \mathcal{O}(\delta^2)
\end{align}
where $\delta \sim \mathcal{N}(0,\sigma^2)$. 
The effect of these results is to change the gate-dependent step lengths contributing to the statistics of the random walk. For this bandwidth of noise, the gate-dependent step lengths become \mbox{$(\tfrac{1}{\sqrt{2}}\hat{n}_1 + \tfrac{1}{\sqrt{2}}\hat{n}_2), (\tfrac{1}{2}\hat{n}_1 +  \tfrac{1}{2}\hat{n}_2), \tfrac{\pi}{2\sqrt{2}}\hat{n}_1$} for $\pi, \tfrac{\pi}{2}, \Identity$ gates respectively with $\hat{n}_1,\hat{n}_2 \in \{\sigx,\sigy,\sigz\}$. The updated expectation value of the distribution is
\begin{equation}
	\E{\noiseAve{ P(\ket{1}) }}\approx J \sigma^2 \tfrac{2}{3}\left(\tfrac{1}{2}+\tfrac{\pi^2}{192}\right). 
	\end{equation}
and the variance for the uncorrelated higher bandwidth error becomes
\begin{align}
\Var{\noiseAve{ P(\ket{1}) }} &= \tfrac{J^2 \sigma^4}{n}\left(\tfrac{4}{9}\left(\tfrac{1}{2}+\tfrac{\pi^2}{192}\right)^2+\tfrac{1}{J}\left(3\left(\tfrac{1}{6}+\tfrac{\pi^4}{2304}\right)\nonumber \right.\right.\\
&\left.\left.-\tfrac{8}{9}\left(\tfrac{1}{2}+\tfrac{\pi^2}{192}\right)^2\right)+\tfrac{(n-1)}{J}\left(\tfrac{1}{6}+\tfrac{\pi^4}{2304}\right.\right.\nonumber \\
&\left.\left.-\tfrac{4}{9}\left(\tfrac{1}{2}+\tfrac{\pi^2}{192}\right)^2\right)\right).
\end{align}
Using these results, we update the approximated Gamma distribution for uncorrelated error processes shown in \eqref{eq:2DUncorOneConGamma} to account for this higher bandwidth noise,
\begin{align}
\langle P(\ket{1})) \rangle_{n,\textrm{uncorrelated}} &\sim \Gamma(\alpha = n, \beta = \tfrac{2}{3n}J\sigma^2  (\tfrac{1}{2}+\tfrac{\pi^2}{192}) ) \label{eq:2DUncorTwoConGamma}.
\end{align}
The renormalized Gamma distributions for uncorrelated error processes shown by solid red lines in the main text Figs. 2C-E were calculated from first principles using \eqref{eq:2DUncorTwoConGamma} with no free parameters. 

To increase this bandwidth for eight noise values in a gate, we study the effect of noise that changes every $\pi/8$ times for a $\pi$ gate and every $\pi/16$ times for a $\pi/2$ gate. In addition, for CORPSE gates, we will also need to consider noise that takes 16 values in an $\Identity$ gate (equivalent to changing every primitive $\pi/16$ time for a primitive $\Identity$, which is executed as a wait equivalent to the length of a $\pi$ pulse).
The error maps to first order in $\delta$ for primitive $\pi$ gates with $\pi/8$ noise, and $\pi/2$ and $\Identity$ gates with $\pi/16$ noise can be calculated in terms of $\delta_{1,\dots,8}, \delta_{1,\dots,16}$. These are rewritten with a single $\delta\sim\mathcal{N}(0,\sigma^2)$ using the Gaussian distributed variable relation \eqref{eq:gaussSum}.
\begin{subequations}
\begin{align}
\Lambda^{(\X)}(\pi,\delta_{1,\dots,8})
&\equiv \mathbb{I}  
- \tfrac{i}{\sqrt{2}} \left\{  \sqrt{ \left(4  - 2\sqrt{2+\sqrt{2}}\right)  }  \right\}  \delta\sigy   
+ \tfrac{i}{\sqrt{2}}  \left\{  \sqrt{ \left(4  - 2\sqrt{2+\sqrt{2}}\right)  }  \right\} \delta\sigz
+  \mathcal{O}(\delta^2) 	 	\nonumber \\
&= \mathbb{I}  
- 0.390 i \delta \sigy 
+ 0.390 i \delta \sigz  
+  \mathcal{O}(\delta^2) 	\\
\Lambda^{(\X)}(\tfrac{\pi}{2},\delta_{1,\dots,8})&= \mathbb{I}  
- 0.196 i \delta \sigy 
+ 0.196 i \delta \sigz  
+  \mathcal{O}(\delta^2) 	\\
\Lambda^{(\Identity)}(\pi,\delta_{1,\dots,16})&= \mathbb{I}  - \tfrac{i\pi}{32} \sum_{i=1}^{16} \delta_i +  \mathcal{O}(\delta^2) \nonumber \\
&\equiv \mathbb{I}  - \tfrac{i\pi}{8} \delta +  \mathcal{O}(\delta^2) 
\end{align}
\label{eq:eightValueErrorMaps}
\end{subequations}
As with the $\pi/2$ uncorrelated noise, the effect of increasing the bandwidth is to change the gate-dependent step contributions to the random walk. From the error maps in \eqref{eq:eightValueErrorMaps}, the step lengths are found to be \mbox{$(0.390 \hat{n}_1 + 0.390 \hat{n}_2), (0.196 \hat{n}_1 +  0.196 \hat{n}_2), \tfrac{\pi}{8}\hat{n}_1$} for $\pi, \tfrac{\pi}{2}, \Identity$ gates respectively with $\hat{n}_1,\hat{n}_2 \in \{\sigx,\sigy,\sigz\}$. 
Finally, before calculating the expectation and variance for CORPSE gates, we need to take into account the relative gate lengths. For primitive gates, an $\Identity$ gate has the same duration as a $\pi$ gate and a $\pi/2$ gate has half the duration, $\tau_\Identity = \tau_\pi,  \tau_{\pi/2}= \tfrac{1}{2}\tau_\pi$. However, due to the $X_{\pi,C}$ followed by  $-X_{\pi,C}$ construction of the CORPSE $\Identity$, it has \emph{twice} the duration as a single CORPSE $\pi$ gate, and a $\pi/2$ has approximately the same duration, $\tau_{\Identity ,C}= 2\tau_{\pi, C},$  $\tau_{\pi/2, C}= 0.92 \tau_{\pi,C}$. As such, to account for their increase in duration relative to a $\pi$ gate, we weight the random walk step contribution from the $\Identity$ and $\pi/2$ gates by a factor of 2 and $8/(13/3) = 1.85$ respectively. 
The updated expectation value of the distribution is
\begin{align}
\E{\noiseAve{ P(\ket{1}) }} &\approx J \sigma^2 
\left( \tfrac{1}{36}(2\times\tfrac{\pi}{8})^2 
+ \tfrac{1}{18} (0.390\sqrt{2})^2 + \tfrac{1}{9} ( 0.390 )^2 + \tfrac{2}{9}  (8/(13/3)\times 0.196\sqrt{2})^2 +  \tfrac{4}{9}  (8/(13/3)\times 0.196)^2   \right)  \nonumber \\
&=  0.167 J \sigma^2 
\end{align}
and the variance for the uncorrelated higher bandwidth error becomes
\begin{align}
\Var{\noiseAve{ P(\ket{1}) }} &= \tfrac{J^2 \sigma^4}{n}
\left( 0.167^2 + \tfrac{1}{J}\left(3\left( 0.041 \right) 
-2(0.167)^2\right)+ \tfrac{(n-1)}{J}  \left(0.041 - 0.167^2\right) \right)   \nonumber \\
&= \tfrac{J^2 \sigma^4}{n} 
 0.028 + \tfrac{0.067}{J}+ 0.013\tfrac{(n-1)}{J}.
\end{align}
\\
%
\subsection{Simultaneous Correlated and Uncorrelated Error Processes}
%


To extract the correlated and uncorrelated error strengths present during execution of a quantum circuit, we combine the previous results examining how the variance of the noise-averaged distribution changes with further noise averaging for different error processes. 
Consider two independent error processes experienced by a quantum circuit with different temporal correlation lengths: one long, $\dL \sim \mathcal{N}(0,\sigLSq)$, and one short, $\dS \sim \mathcal{N}(0,\sigSSq)$. The first process is taken to be maximally correlated across the length of a circuit, with block length $\Mn = J$, whilst the second varies randomly every primitive $\pi/2$ time. This results in two simultaneous random walks in Pauli space, $\RvecL = \dL \VvecL$ and $\RvecS$. We expand the expression for survival probability using the 2D projection of these vectors in the $\sigx \- \sigy$-plane in Pauli space, $\RvecLXY, \RvecSXY$,
\begin{align}
1- \noiseAve{ P(\ket{1}) } &= \noiseAve{ \norm{\RvecSXY + \dL\VvecLXY}^2  } \nonumber\\
&= \noiseAve{ \RnormSQSXY } + \noiseAve{ \dL^2 \VnormSQLXY } + \noiseAve{ 2\dL \RvecSXY \cdot \VvecLXY } \nonumber   \\
&=  \noiseAve{\RnormSQSXY} + \sigLSq \VnormSQLXY    
\end{align} 
using $\noiseAve{\dL} = 0$ for $\dL\sim\mathcal{N}(0,\sigLSq)$. Then, the variance is
\begin{align}
\Var{1- \noiseAve{ P(\ket{1}) }} &= \Var{  \noiseAve{\RnormSQSXY} + \sigLSq \VnormSQLXY   } \nonumber \\
&= \Var{  \noiseAve{\RnormSQSXY} } + \sigLFour \Var { \VnormSQLXY} + 2\sigLSq\Cov{\noiseAve{\RnormSQSXY} ,  \VnormSQLXY}.
\end{align} 
For primitive gates (no scaling of gate lengths), the expression for error variance scaling under simultaneous error processes becomes
\begin{align}
\Var{\noiseAve{ P(\ket{1}) }} &= \left\{ \tfrac{J^2 \sigSFour}{n}\left( \tfrac{4}{9}\left(\tfrac{1}{2}+\tfrac{\pi^2}{192}\right)^2+\tfrac{1}{J}\left(3\left(\tfrac{1}{6}+\tfrac{\pi^4}{2304}\right) \nonumber  \right.\right.\right. \\
&\left. -\tfrac{8}{9}\left(\tfrac{1}{2}+\tfrac{\pi^2}{192}\right)^2\right)+\tfrac{(n-1)}{J}\left(\tfrac{1}{6}+\tfrac{\pi^4}{2304} \right.\nonumber \\
&\left.\left.\left. -\tfrac{4}{9}\left(\tfrac{1}{2}+\tfrac{\pi^2}{192}\right)^2\right)\right) \right\}   \\
&+  \left\{   \tfrac{J^2 \sigLFour}{n}\left(\tfrac{12}{9}\left(\tfrac{1}{2}+\tfrac{\pi^2}{96}\right)^2+\tfrac{1}{J}\left(3\left(\tfrac{7}{36}+\tfrac{\pi^4}{576}\right)\nonumber \right.\right.\right. \\
&\left. -\tfrac{8}{3}\left(\tfrac{1}{2}+\tfrac{\pi^2}{96}\right)^2\right)+(n-1)\left(\tfrac{4}{9} \left(\tfrac{1}{2}+\tfrac{\pi^2}{96}\right)^2+\nonumber \right.\\
&\left.\left.\left. \tfrac{1}{J}\left(\tfrac{7}{36}+\tfrac{\pi^4}{576}-\tfrac{8}{9}\left(\tfrac{1}{2}+\tfrac{\pi^2}{96}\right)^2\right)\right)\right)   \right\} \\
&+ \left\{ 2J \sigLSq \sigSSq ( (\tfrac{1}{6} + \tfrac{\pi^4}{1152}) - \tfrac{4}{9}(\tfrac{1}{2}+\tfrac{\pi^2}{96})(\tfrac{1}{2}+\tfrac{\pi^2}{192})) \right\}.
\end{align}
For CORPSE gates, we combine 8$\times$ uncorrelated noise bandwidth calculations above with the original primitive correlated calculations, where the relative detuning contributions have been scaled by $1, 2$ or $8/(13/3)$ for $\pi$, $\Identity$ and $\pi/2$ gates respectively, yielding 
\begin{align}
\Var{\noiseAve{ P(\ket{1}) }} &= \left\{ \frac{J^2 \sigSFour}{n} \left( 0.028 + \frac{0.067}{J}+ 0.013\frac{(n-1)}{J} \right) \right\}  \nonumber \\
&+  \left\{   \frac{J^2 \sigLFour}{n} \left(3\times1.14^2+\frac{1}{J}(3\times 3.78 - 6\times1.14^2)+(n-1)(1.14^2+ \frac{1}{J}(3.78 - 2\times1.14)) \right)   \right\}  \nonumber \\
&+ \left\{ 2J \sigLSq \sigSSq \left( 0.318 -1.142\times0.167 \right) \right\}.
\end{align}
Fitting this result to the mean variance trajectories obtained in the main text found $\sigSSq =7.3\times10^{-3}$, $\sigLSq = 5.6 \times 10^{-6}  $ for CORPSE gates and $\sigSSq = 8.6\times10^{-3}$, $\sigLSq =   1.3 \times 10^{-4}$ for WAMF. For the WAMF, we scale the relative detuning contributions by $1, 2$ or $1.57$ for $\pi$, $\Identity$ and $\pi/2$ gates respectively.
Comparing the extracted error to the applied noise strengths \mbox{$\sigSSq = 5.2\times10^{-4}$, $\sigLSq =   2.1 \times 10^{-3} $}, we find a $370\times$ suppression in the correlated component from CORPSE and $16\times$ from WAMF. Even taking the largest value for $\sigLSq$ after applying CORPSE from the confidence intervals calculated in Section~\ref{sec:statistical_tests} shows a suppression of $254\times$.

%
\newpage
\section{Influence of Quantum Projection Noise}
%


Quantum projection noise (QPN) describes the intrinsic uncertainty in qubit measurements due to the binomial nature of quantum state collapse \cite{Itano:1993} and its scaling with the number of samples. The variance of a measurement due to QPN is $\nicefrac{p(1-p)}{r}$, where p is the true state projection along the $z$-axis of the Bloch sphere and $r$ is the number of identical measurements performed. Our work studies variances over distributions of noise-averaged survival probabilities, and consequently it is necessary to demonstrate that we were not limited by QPN bounds. In order to ensure that our results are not measurement artefacts from quantum projection noise, we average each circuit and noise realization combination $r=220$ times. At this number of repetitions, the largest possible projection noise variance is given by $\nicefrac{0.5(1-0.5)}{220} = 1\times10^{-3}$. 

In addition to the worst case QPN, we compare the variance scaling results for the CORPSE DCG under simulataneously applied correlated and uncorrelated noise to the QPN given by the measured survival probabilities. Fig.~\ref{SuppFig:QPN} shows the mean trajectory for the CORPSE variance scaling under the combined noise process presented in main text Fig.~3C in dark blue. The dashed black line gives the worst case QPN and the two other sets of trajectories are calculated directly from the measured probabilities. For these, the QPN was calculated at each $n$ for 100 randomizations of noise realizations to reduce bias and the 100 values are plotted. The lower set of trajectories are divided by $(n\times r)$ rather than just $r$. Our results are well above this lower limit suggesting that this is the most valid measurement of setting our QPN limit. Furthermore, we note that the saturation observed at large values of $n$ is not set by any static QPN bound limiting our measurements.  
\begin{figure}[h]
\centering
\includegraphics[scale = 1]{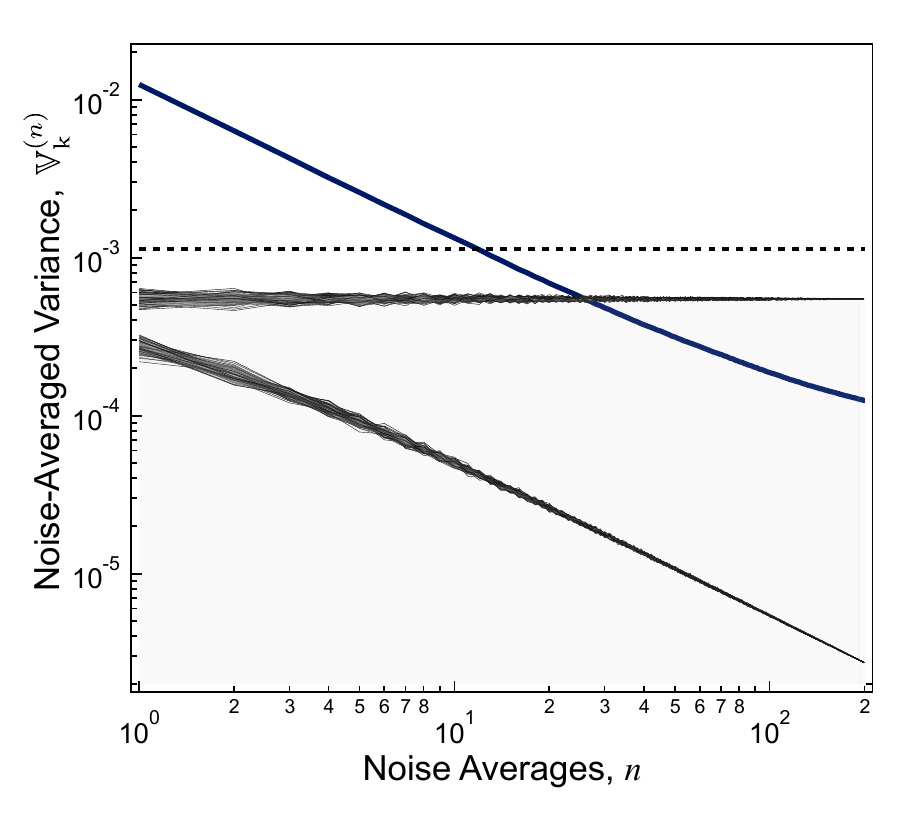}
\caption{\textbf{Quantum projection noise limits for measured survival probabilities with the CORPSE DCG} 
Comparison of mean CORPSE variance scaling from main text Fig. 3C to QPN variance limits given by $p(1-p)/r$. Dashed line is worst case QPN for $r=220$ when $p = 0.5$. Black lines show additional QPN limits where, for each $n$, $p(1-p)/r$ is calculated for 100 randomizations of noise realizations. The lower line scaling as $1/n$ is divided by $(n\times r)$ rather than $r$.
}
\label{SuppFig:QPN}
\end{figure}

%
\newpage
\section{Statistical Analysis of Extracted Error Strengths}
\label{sec:statistical_tests}
%


In this work, we attempt to quantify the performance of CORPSE DCGs to suppress correlated error, claiming a factor of $\sim260\times$ suppression relative to the applied correlated noise strength and the correlated error strength that was experienced by primitive gates under the same applied noise. This is based off the analytic model developed from the random walk framework in \cite{Ball:2016}, which has been modified appropriately for our experimental framework. After applying noise with correlated component strength $\sigLSq = 1.986\times10^{-3}$ and uncorrelated component strength $\sigSSq = 0.517\times10^{-3}$, we extract corresponding error strengths of $\sigLSq = 5.6\times10^{-6}$ and $\sigSSq = 7.2\times10^{-3}$ respectively.

We use the Akaike Information Criterion (AIC) \cite{Akaike:1974} to test how good the model of $\sigLSq = 5.6\times10^{-6}$ is within the analytic framework provided. This is done by allowing $\sigSSq$ to vary freely whilst $\sigLSq$ is fixed at values increasing from 0, and the AIC is calculated using the maximum likelihood estimate, $\textrm{RSS}/n$, where RSS is the Residual Sum of Squares from the model. The AIC is given by
\begin{equation}
\textrm{AIC} = 2k + n\textrm{ln}(\textrm{RSS})
\end{equation}
where the number of estimated parameters is $k=2$: $\sigLSq, \sigSSq$. From this, we can calculate the relative likelihood of each possible model $i$ using
\begin{equation}
\textrm{AIC}_\textrm{Rel} = \textrm{exp}((\textrm{AIC}_\textrm{Min} - \textrm{AIC}_i)/2).
\end{equation}
The relative likelihood is shown in Fig. \ref{SuppFig:AICBIC}A, and we find a 95\% likelihood for $\sigLSq = (5.6\substack{+1.9 \\ -2.3})\times10^{-6}$. 

The Bayesian Information Criterion (BIC) \cite{Schwarz:1978} can be derived from the same framework as the AIC, but with an alternative prior. It is calculated in a similar manner,
\begin{equation}
\textrm{BIC} = \textrm{ln}(n)k  + n\textrm{ln}(\textrm{RSS/n})
\end{equation}
and shows strong model violation when $\Delta\textrm{BIC} \coloneqq \textrm{BIC} - \textrm{BIC}_\textrm{Min} > 10$. Here, this occurs outside the range $\sigLSq = (5.6\substack{+2.6 \\ -3.2})\times10^{-6}$, as shown in Fig. \ref{SuppFig:AICBIC}B.
\begin{figure}[h]
\centering
\includegraphics[scale = 1]{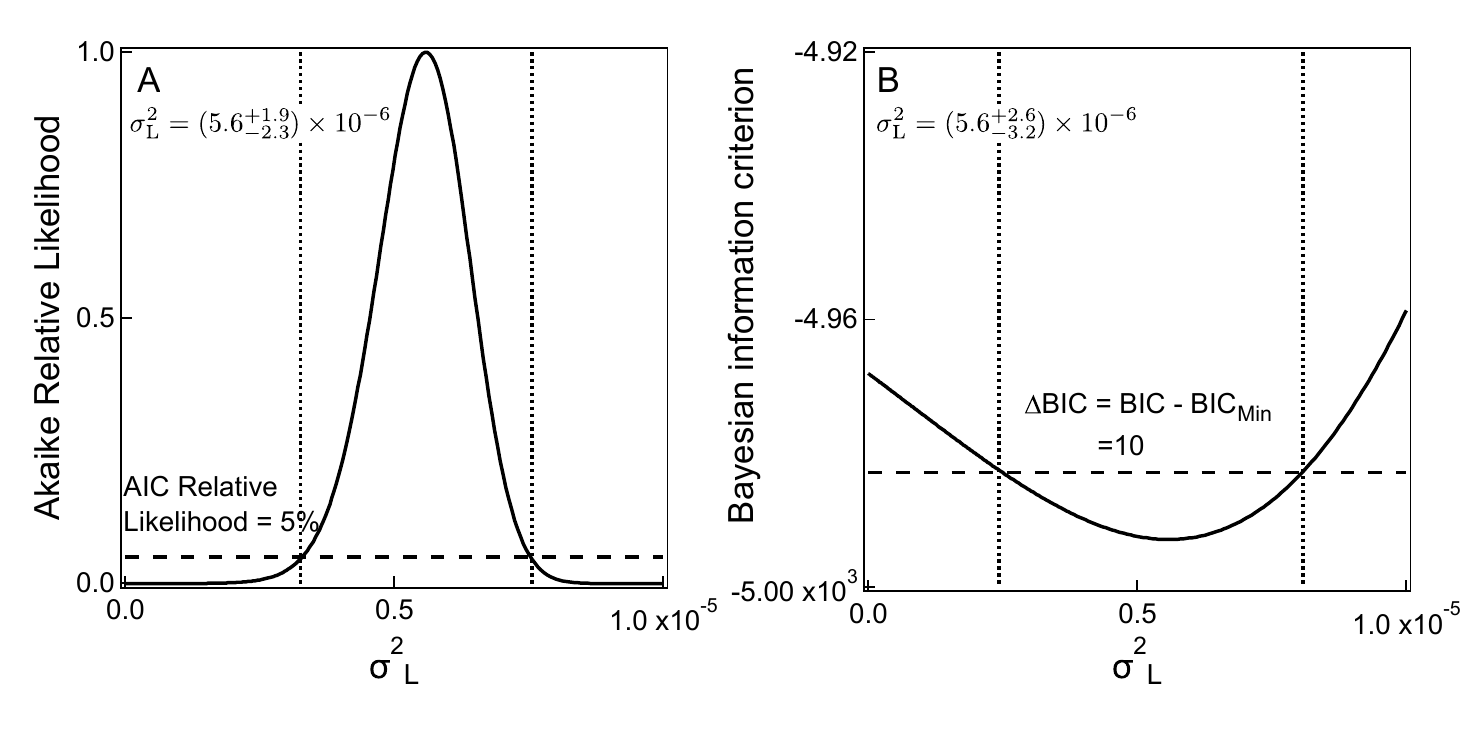}
\caption{\textbf{Statistical analysis of extracted correlated error strength for CORPSE DCG} 
(A) Relative likelihood derived from the Akaike Information Criterion (AIC) with $k=2$ free parameters, shows that varying $\sigLSq$ gives a 95\% likelihood bound within the range $\sigLSq =5.6\protect\substack{+1.9 \\  -2.3}  \times 10^{-6}$ when applying the model presented in this work. The dashed line shows the 5\% relative likelihood cutoff, such that all values of $\sigLSq$ within the dotted lines have $\geq95\%$ likelihood. (B) Similarly, the Bayesian Information Criterion (BIC) shows strong model violation ($\Delta\textrm{BIC} > 10$) for \mbox{$\sigLSq = (5.6\protect\substack{+2.6 \\ -3.2})\times10^{-6}$}. The dashed line indicates the strong model violation cutoff, with dotted lines showing the corresponding bounds of $\sigLSq$.}
\label{SuppFig:AICBIC}
\end{figure}

\clearpage
\end{widetext}
\end{document}